\documentclass[10pt,twocolumn,twoside]{article}
\usepackage{epsf,latexsym}
\begin{document}
\title{(Anti)proton and Pion
Source Sizes and Phase Space Densities in Heavy Ion Collisions}
\author{Michael Murray\\
 Cyclotron Institute, Texas A\&M University, College Station, TX 77843-3366}
%\abstract{
\maketitle
\begin{abstract}
NA44 has  measured mid-rapidity deuteron spectra from
$AA$ collisions at $\sqrt{s_{nn}}\approx 18$ GeV at the CERN SPS.
Combining these spectra with published $p$, $\bar p$ and $\bar d$ 
data 
allows us to calculate, within a coalescence framework, $p$ and $\bar p$
source sizes and phase space densities.
These results are compared to $\pi^\pm$ source sizes
measured by Hanbury-Brown Twiss, HBT, interferometry and
phase densites produced by combining pion spectra and HBT results.
We also compare to
$pA$ results and to lower energy (AGS) data.
 The $\bar p$ source is larger than
the proton source at  $\sqrt{s_{nn}}=17.3$ GeV.
The phase space densities of $\pi^+$ and $p$ are not constant but 
grow with system size. Both $\pi^+$ and proton radii decrease with 
$m_T$ and increase with $\sqrt{s_{nn}}$.
Pions and protons do not freeze-out independently. The nature of 
their interaction changes as $\sqrt{s_{nn}}$ and the $\pi/p$ ratio 
increases.
%\pacs{PACS numbers: 25.75.-q 25.75.-gz 25.75.-ld 13.85.-t 25.40.ve
% Version 9.1 \today}
\end{abstract}

Relativistic heavy ion collisions provide a mechanism
to heat and compress nuclear matter to 
 temperatures and energy densities comparable to those of 
the early universe when it was still a 
plasma of quarks and gluons.
 Such a state may be fleetingly restored in 
these collisions 
where temperatures of $ T= 168 \pm 3$ MeV and 
energy densities $\epsilon = 3$ GeV/fm$^3$
are observed \cite{PBM99A,NA49ET}. 
These values are close to those of the phase transition 
found in lattice calculations \cite{LQCD}. 
If such a hot and dense state were formed one would expect 
a large increase in entropy and possibly a saturation of the
density of particles in phase space. 
The coalescence of nucleons into deuterons is sensitive to 
both their spatial and 
momentum correlations. 
In this paper we report $p$, $\bar p$ and $\pi^+$ 
source sizes measured by coalescence and interferometry, and 
combine these with single particle spectra to derive 
phase space densities.
The phase space densities depend on
temperature, chemical potentials, and velocity fields in the system.
This description of the final hadronic state serves as a boundary 
condition 
for models of possible quark gluon plasma production.
We vary the total size of the system by studying 
$PbPb$, $SPb$, $SS$ and $pPb$ collisions. 
We also
compare our results to lower energy AGS data where the $\pi/p$ ratio
is much lower.  This comparison shows that
the freeze-out of pions and protons is coupled.

\begin{figure}
 \mbox{\epsfxsize=8.0cm\epsffile{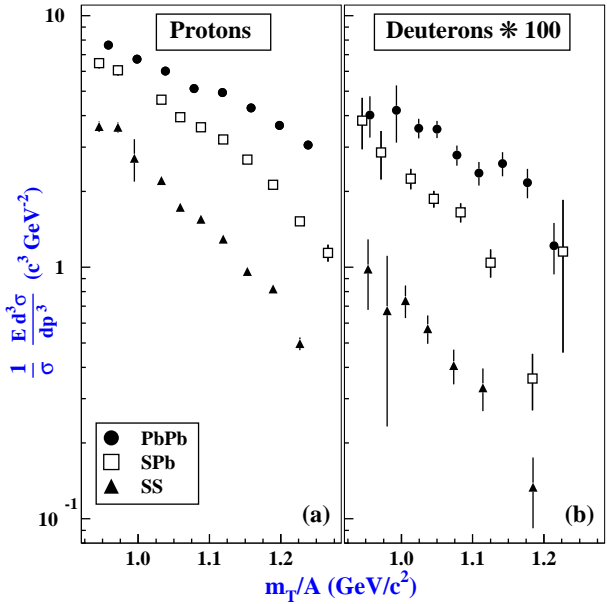}}
 \caption{Invariant cross-sections vs. $m_T/$A at y=1.9-2.3, for protons (a)
and deuterons (b) from central $SS,SPb$ and $PbPb$ collisions
\protect{\cite{NA44pmp,thankN44}}}
 \label{fg:dpvmta}
\end{figure}

NA44 is a focusing spectrometer that uses
conventional dipole magnets and superconducting quadrupoles to
analyze the momentum of the produced particles and create a magnified 
image of the target in the spectrometer 
\cite{NA44,NA44pbk,NA44pbpi,NA44spp,NA44rvm,NA44pmp,NA44dbar}.
 The systematic errors on the deuteron
yields range from $14\%$ for $SPb$ to $9\%$ for $PbPb$.
The $p$ and $\bar p$ spectra
are corrected for feed-down from $\Lambda$ and $\Sigma$ decays using
a GEANT simulation with the ($\Lambda/p$) and ($\Sigma/p$) ratios taken
from the RQMD model \cite{NA44pmp,NA44dbar,SORGE95}.
 The systematic error was estimated by varying these ratios
by $\pm25\%$.
These errors are slightly correlated for $p$ and $\bar p$.
Fig.~\ref{fg:dpvmta} shows NA44
 deuteron spectra  and previously published proton spectra at
$y$ = 1.9-2.3 as a function of $m_T/$A
  \cite{thankN44}.  The centrality
is $\approx 10\%$ for $SS$, $SPb$ and $PbPb$. The spectra get flatter for
the larger systems consistent with a higher temperture and/or stronger sidewards
 flow.
As expected from coalescence, the slopes
are similar for protons and deuterons. This  was also found for lower energy data,
\cite{E866pd}.
The deuteron inverse slopes (in $m_T$) and yields are listed in
Table \ref{tb:dslope}.
 A  comprehensive
analysis of
all NA44's  proton and light clusters spectra will be given in a later
paper.

\begin{table}
\centering
  \begin{tabular}{|c|r|l|r|} \hline
{\bf System} & \multicolumn{1}{c|}{\bf Fit Range} &
\multicolumn{1}{c|}{\bf Inverse} & \multicolumn{1}{c|}{\bf dN/dy} \\
 & \multicolumn{1}{c|}{\bf of $m_T-m$ }& \multicolumn{1}{c|}{\bf Slope}  & \\
 & \multicolumn{1}{c|}{$MeV/c^2$} & \multicolumn{1}{c|}{$MeV/c^2$}  &
 \multicolumn{1}{c|}{($10^{-2}$)} \\ \hline
 SS   & 0-520     & 320 $\pm$ 149  &  39 $\pm$ 13 \\
 SPb  & 0-520     & 300 $\pm$ ~~91  & 153 $\pm$ 23 \\
 PbPb & 160-520   & 379 $\pm$ ~~13  & 390 $\pm$ 20 \\ \hline
  \end{tabular}
  \caption{Deuteron inverse slopes and yields.
Systematic and statistical errors have been added in quadrature.
The errors are dominated by statistics and the extrapolation out of the
acceptance. The PbPb fit is from \protect{\cite{HANSENQM}.}}
  \label{tb:dslope}
\end{table}

The model of deuteron production by final state coalescence
of protons and neutrons with small relative momenta 
states that the production of deuterons with a certain velocity is
proportional to the number of protons and neutrons that have similar
velocities \cite{BUTLER,KAPUSTA}. 
This 
model successfully describes measured deuteron distributions in intermediate
energy heavy ion collisions and high energy $pA$ collisions, 
\cite{COALDATA}. Near mid-rapidity, direct 
production of $d\bar d$ pairs is small 
due to the high $d\bar d$ mass threshold of 3.75 GeV/{\it $c^2$}, and
pre-existing
deuterons are unlikely to survive the many collisions required
to shift them 
to mid-rapidity.
Since coalescence depends on the distribution of nucleons,
one can determine a nucleon source size from 
the  ratio
 \begin{equation}
 B_2(p)
\equiv
 \frac{ \frac{E_{p} d^3N_{p}}{dp^3}
  \frac{E_{n} d^3N_{n}}{dp^3}}{\frac{E_{d} d^3N_{d}}{dP^3}}
 \label{eq:b2}
\end{equation}
where the deuteron momentum $P$ is twice the  proton momentum $p$ \cite{MEKJIAN}.
 Since we do not measure neutrons we assume
that  the spectra have the same shape and take the $n/p$ ratio to be
$1.06\pm.04$ from RQMD \cite{SORGE95}. At $\sqrt{s_{nn}}=4.9$ GeV
the measured $n/p$ ratio is $1.19\pm.08$ independent of $m_T$
 \cite{E864Neut}. RQMD is in reasonable agreement with this result.

\begin{figure}%[h]
 \mbox{\epsfxsize=8cm\epsffile{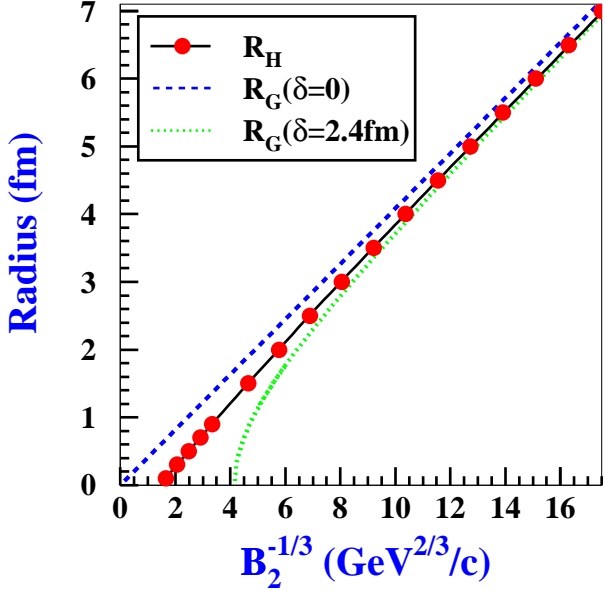}}
 \caption{Comparison of proton source radii versus the coalesence parameter
$B_2^{-1/3}.$  $R_H$, shown by $\bullet$, is derived numerically using the Hulthen wave function in
 Eqn.\protect{\ref{eq:wavefunc}}.
 The solid line is a fit to the numerical results.
The dotted and dashed curves
show $R_G$ from
   Eqn.\protect{\ref{eq:rg}}
 with and without a correction for the deuteron size.}
 \label{fg:r0rgb2}
\end{figure}

To facilitate comparison with NA44's pion interferometry results, 
we assume a Gaussian distribution of 
the proton source. If one also assumes a Gaussian wave-function
one can solve for the source size analytically \cite{SCOTT95}
\begin{equation}
 (R_G^2+\frac{\delta^2}{2})^{3/2} = \frac{3\pi^{\frac{3}{2}}(\hbar c)^3}{2m_p B_2}
 \label{eq:rg}
\end{equation}
where $m_p$ is the proton mass
and $\delta=2.8$ fm accounts for the
size of the deuteron. In reality the deuteron wave function is not
Gaussian but is more accurately represented by the Hulthen form
\begin{equation}
\phi(r)=\sqrt{\frac{\alpha\beta(\alpha + \beta)}{2\pi(\alpha - \beta)}}  \cdot
\frac{e^{-\alpha r}- e^{-\beta r}}{r}
\label{eq:wavefunc}
\end{equation}
with $\alpha=0.23$ fm$^{-1}$ and $\beta=1.61$ fm$^{-1}$  \cite{HODGSON}.
The convolution of such a wave function with a gaussian source cannot be done
analytically but is  straightforward numerically to solve for the source
radius $R_H$, \cite{MROWCZ92}. Note our $R_H$ is the $R_0$ of   \cite{MROWCZ92}.
Figure~\ref{fg:r0rgb2} shows a comparison of $R_H$ and $R_G$ versus the
coalesence parameter $B_2$.

Since $R_H$ is sensitive to both the transverse and longitudinal size
of the source, when comparing to HBT results  it is best to compare 
to $(R_\perp^2 \cdot R_\parallel)^{\frac{1}{3}}$ (Eqn. 6.3 of \cite{SCHEIBL}), 
where $R_\perp$ and $R_\parallel$ parametrize the extent of the source 
perpendicular and parallel to the beam \cite{HEINZ96}. NA44 has
published HBT results in the Pratt-Bertsch frame in which the sum of the
longitudinal momentum of each pion pair is zero. In this scheme the radius in
the sidewards direction
$R_s = R_\perp$ and the longitudinal radius
$R_l \approx R_\parallel$ \cite{CHAPMAN}.
 In this paper
we will therefore compare $R_H$ to
$(R_s^2 \cdot R_{l})^{\frac{1}{3}}$.
These 
parameters can be thought of as 
``lengths of homogeneity'' of the source \cite{CSOR94A,WEID99A}.
One can think of the radii as lengths scales of the velocity and/or temperature
 gradients.
 They  represent snapshots of the hadronic
system at
freeze-out which may occur at different times for $\pi^+$ and protons. 
However since the cross sections for $\pi\pi$, $p\pi$ and $pp$
collisions are comparable the freeze-out times should be close.

A particle's phase space density is defined as 
\begin{equation}
 f({\bf p}, {\bf x}) \equiv (2\pi \hbar c)^3
  \frac{d^6N}{dp^3dx^3}.
 \label{eq:fdefine}
\end{equation}
For a system in chemical  equilibrium at a temperature  $T$ and chemical potential $\mu$
\begin{equation}
 f(E) = \frac{1}{e^{(E-\mu)/T}\pm 1} 
 \label{eq:fbose}
\end{equation}
where E is the energy and $\pm 1$ selects bosons or fermions.
For a dilute system, {\it i.e.} $f \ll 1$,
Eqn.~\ref{eq:fbose} gives
%\begin{eqnarray}
% f_p & \approx & e^{-(E_p-\mu )/T} \\
% f_n & \approx & e^{-(E_n-\mu )/T} \\
% f_d & \approx & e^{-(E_d-2\mu )/T}.
% \label{eq:fpnd}
%\end{eqnarray}
\begin{equation}
 f_d  \approx  e^{-(E_d-mu_p-\mu_n )/T}.
 \label{eq:fpnd}
\end{equation}
%Where $p,n$ and $d$ represent protons, neutrons and deuterons.
Since $E_d = E_n + E_p$, Eqn.~\ref{eq:fpnd} implies that
\begin{equation}
 f_d({\bf P}, {\bf x})=f_p({\bf p}, {\bf x}) f_n({\bf p}, {\bf x}) .=
\frac{n}{p} \cdot f_p({\bf p}, {\bf x})^2
\end{equation}
A more general form of this relation was derived in Eqn.~3 of 
\cite{POLL98A} assuming only that the system is hot and large compared
to the deuteron. 
Averaging $f_p$ over ${\bf x}$ gives 
\begin{equation}
 \langle f_p \rangle = \frac{1}{3}
 \frac{E_d\frac{d^3N_d}{dP^3}}{E_p\frac{d^3N_p}{dp^3}} \cdot \frac{p}{n}
 \label{eq:pfaze}
\end{equation}
where the factor of 3 accounts for the spin of the particles.
For pions NA44 has measured the source size in
3 dimensions with HBT, 
as well as single particle 
spectra.
Some of the pions come from the
 decay of long-lived resonances such as $\eta, \eta'$ and $\omega$. 
These pions reduce the strength of the correlation function $\lambda$,
which typically is $<1$. The fraction of pions which do not
come from resonances is $\sqrt{\lambda}$, \cite{CSOR96A}.
This has been shown experimentally for $e^+e^-$ collisions and for 
RQMD simulations of 
$PbPb$ collisions \cite{HUBERT,JOHNQM97}. 
Dividing $\sqrt{\lambda} d^3N_\pi/dp^3$ by the Lorentz invariant volume, 
\cite{WEID99A,BERTSCH,E87797}
gives
%\sqrt{\lambda} {d^3N_\pi\over dp^3} \frac{1}{R_s\sqrt{R^2_o R^2_l - R^4_{ol}}} .
\begin{equation}
 \langle f_\pi \rangle  =\frac{\pi^{\frac{3}{2}} (\hbar c)^3}{3}
\sqrt{\lambda} {d^3N_\pi\over dp^3} \frac{1}{R_{s}\sqrt{R^2_{o}
R^2_{long} - R^4_{ol}}} .
 \label{eq:pifaze}
\end{equation}
Again the factor of 3 accounts for the pion's spin degeneracy.
Here $R_l$ is the extent of the source along the beam direction;
$R_o$ the extent in the outward direction, ie towards the observer and
$R_s$ measures the source in the sidewards direction, perpendicular
to both the beam axis and the line of sight to the observer. The
$R_{ol}$ term is the ``out-longitudinal" cross term \cite{CHAPMAN}.
For $PbPb$ collisions setting $R_{ol}=0$ in the HBT fit increases 
$\langle f_\pi \rangle$ by $9\% \pm 10\%$, \cite{NA44pbpi}. 
For $pPb$, $SS$ and $SPb$  we assume $R_{ol}=0$ 
but add a systematic error of $13\%$  to
 $\langle f_\pi \rangle$.
For $pPb$ 
deuteron
spectra are not available and $\langle f_p \rangle$
 was calculated using Eqn.~\ref{eq:pifaze}, replacing
the last term with $\frac{1}{R_{inv}^3}$. 
$R_{inv}$ was determined from $pp$  HBT 
data \cite{NA44spp}.
%This is justified since for
For $PbPb$ $R_H$ and $R_{inv}$ agree within their errors
of $\approx 5\%$, \cite{MURRAYQM}.

\begin{figure}%[h]
 \mbox{\epsfxsize=8.5cm\epsffile{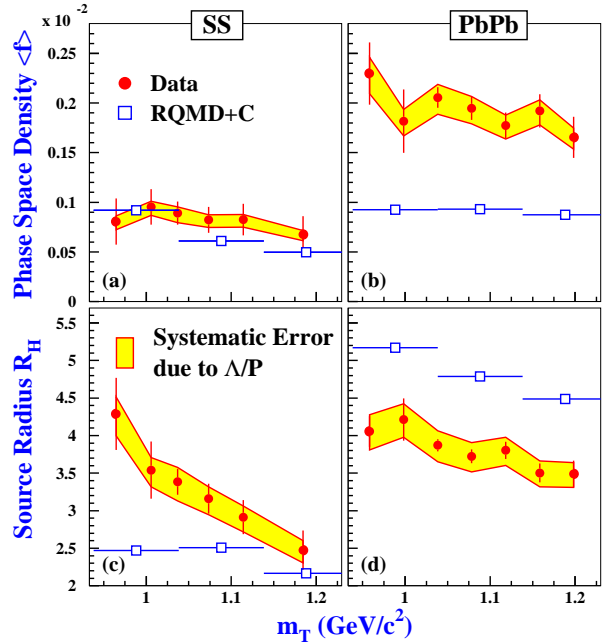}}
 \caption{Proton phase space densities $\langle f_p \rangle$ (a),(b)
 and radii $R_H$ (c),(d)
 versus $m_T$ for $\protect{\sqrt{s_{nn}}}= 17.3$ GeV for data $\bullet$
and RQMD $\Box$.
 The shaded bands indicate the estimated systematic error on the correction
 for weak decays.}
 \label{fg:aarfaze}
\end{figure}

We can test the usefulness of these coalesence methods using RQMD,
coupled with a coalesence afterburner, \cite{SORGE95B}.
Figure~\ref{fg:aarfaze} shows a comparison of
$\langle f_p \rangle$
and $R_H$ for both data and RQMD for $SS$ and $PbPb$. For the data
both $\langle f_p \rangle$ and $R_H$ are larger in $PbPb$ but for
RQMD only $R_H$ increases from $SS$ and $PbPb$ while
$\langle f_p \rangle$ stays constant. This invariance of $\langle f_p \rangle$
may be an artifact of the coalesence mechanism used, which ignores the
requirement of a third
particle.
 Since
$\langle f_p \rangle$ is constant in the model the increase in proton
multiplicity from $SS$ to $PbPb$ is accommodated by a large increase in
$R_H$. However, $R_H$ is less than the average transverse radius of
freezeout, 5.4 fm for $SS$ and 10.3 fm for $PbPb$, indicating that coalesence is
not sensitve to the full size of the source. A similar situation
occurs in pion interferometry where correlations between position
and momentum cause the observer to only ``see" the side of the
source closest to her, \cite{NA44pbpi}. Since these correlations
get stronger as the particles get faster the size of the source drops
with $m_T$, \cite{CSOR96A,SINUKOV94}.

\begin{figure}[t]
 \mbox{\epsfysize=10.5cm\epsffile{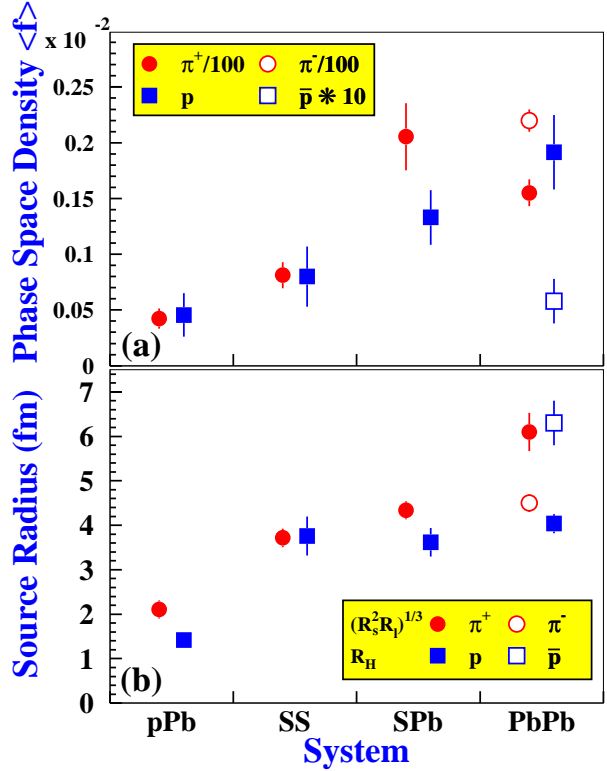}}
 \caption{Phase space densities $ \langle f \rangle $ and
source radii
 for $\pi^+$ and $p$ at $\langle p_T \rangle \approx 240$ MeV/{\it c},
 and for $\bar p$ at  $\langle p_T \rangle \approx 490$ MeV/{\it c}.
 For $pPb$, $\protect{\sqrt{s_{nn}}}=30$ GeV and the proton points
 are derived from $pp$ HBT data
\protect{\cite{NA44spp}}.}
 \label{fg:frsys}
\end{figure}

Figure~\ref{fg:frsys} shows the system dependence of the
phase space densities and source radii for $\pi^{\pm}$, $p$ and $\bar p$.
The $p$ and $\bar p$ radii for $PbPb$ are consistent with coalescence
data at
$p_T=0$, and $pp$ interferometry results \cite{NA52dbar,NA49pbpp}.
The $\pi^+$ and $p$ phase space densities generally
increase with system size.
We find that
\[ \langle f_{\bar p} \rangle \ll \langle f_p \rangle \ll
\langle f_{\pi^+} \rangle < \langle f_{\pi^-} \rangle \ll 1\]
For $SPb$ and $PbPb$ $\langle f_{\pi^+} \rangle$ was calculated in
\cite{FERENC} using a similar equation to Eqn.~\ref{eq:pifaze}. However
a parametrization of the pion spectrum was used rather than the spectrum itself.
The authors of \cite{FERENC} concluded that the pion phase density was
universal at freeze-out but this is not the case,
since $\langle f_{\pi^+} \rangle$ is considerably smaller for $pPb$ and
$SS$ than for $SPb$ and $PbPb$.

For pions $(R_s^2 \cdot R_l)^{\frac{1}{3}}$ increases steadily
with the number of participants and for $PbPb$ there is a rapid increase in
the radii parameters with multiplicity, \cite{NA44rvm}. At low $p_T$
$R_H$ does not change much from $SS$ to $PbPb$ (nor with centrality for $PbPb$)
despite the increase in the proton multiplicity
by  $\approx 3$, see Fig.~\ref{fg:dpvmta}(a). However the
$m_T$ dependence of  $R_H$  is weaker for $PbPb$ than for $SS$,
see Fig~\ref{fg:aarfaze}.
The extra protons mainly increase the proton phase space density $\langle f_p \rangle$.

Because of their large annihilation cross-section, one
might expect that antiprotons (particularly those at low $p_T$)
would be emitted only from the surface of the system and would
have a larger RMS freeze-out radius than protons.
Our data are consistent with this idea.
 An alternative view
assumes that  proton and antiprotons are produced in the same volume
but that
antiprotons are suppressed in the interior of the source \cite{MROWCZ93}.
Applying this
idea to our data would imply that antiprotons are emitted only from
within $1.0\pm0.2$ fm of the surface \cite{MROWCZ93}.
Recent $AuPt$ and $AuPb$
results from E864
at $\sqrt{s_{nn}}=4.9$ GeV and low $p_T$ imply that
$R_H^p=(4.0\pm 0.2)$ fm,
$R_H^{\bar p}=(2.2\pm 0.9\pm 0.6)$ fm and
$\langle f_{\bar p} \rangle=(4.0\pm1.9)\cdot 10^{-6}$
 \cite{E864dbar}.
At $\sqrt{s_{nn}}=4.9$ GeV antiprotons are mainly produced in primary
nucleon-nucleon
collisions and so they may have a smaller source size than protons.
Since $R_H$ is a gaussian radius it is necessary to multiply it by
$\sqrt{5}$ in order to compare it to a hard sphere with the same RMS 
radius, \cite{JACAK95}. If this is done the antiproton source 
 is roughly equal to the size of the colliding $Au$ nuclei at
$\sqrt{s_{nn}}=4.9$ GeV.

\begin{figure}%[h]
 \mbox{\epsfysize=13.7cm\epsffile{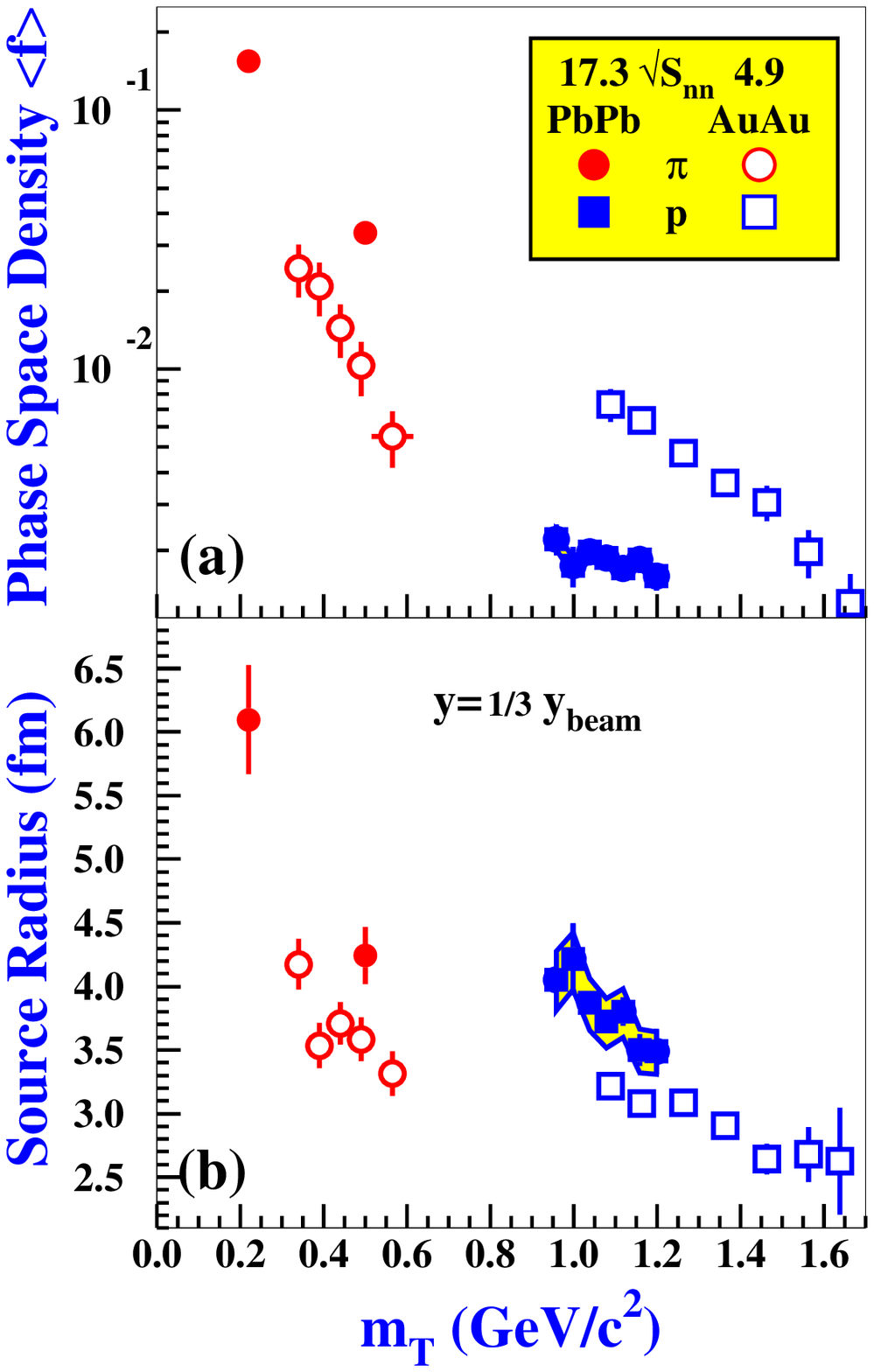}}
 \caption{Phase space densities (a) and radii (b) for $\pi^+$ and $p$
 versus $m_T$ for $\protect{\sqrt{s_{nn}}}$ = 17.3 and $4.9$ GeV.
 The shaded bands indicate the estimated systematic error on the correction
 for weak decays.
 The E866 points used data from \protect{\cite{E866pd,E866jh,E802AHLE}}.}
 \label{fg:hbtdpp}
\end{figure}

In order to study the energy dependence of freeze-out we compare our
$PbPb$ data at $\sqrt{s_{nn}} = 17.3$ GeV to AGS $AuAu$ data at
$\sqrt{s_{nn}}=4.9$ GeV. Since our data are not at mid-rapidity but at y=2
we have compared results at the same scaled rapidity $y=\frac{1}{3}y_{beam}$.
Figure~\ref{fg:hbtdpp} shows the phase space densities and source radii 
for $PbPb$ and $AuAu$ collisions as a function of $m_T$.
At a given $m_T$, $\langle f_{\pi^+} \rangle $ 
increases with $\sqrt{s_{nn}}$ %, in contradiction to \cite{FERENC}, 
while $\langle f_p \rangle$ decreases. 
Fitting $\langle f_{\pi^+} \rangle$ to Eqn.~\ref{eq:fbose} 
gives $\mu_\pi^+=0$, within errors, for both energies
while $\mu_p/T$ decreases with $\sqrt{s_{nn}}$. 
Since $E=m_T cosh(y)$, Eqn.~\ref{eq:fbose} also implies that $f$ be 
exponential in $m_T$ for $f \ll 1$. 
However if the system is boosted due to transverse flow, 
$f(m_T)$ will become flatter \cite{TOMASIK}. 
This effect is proportional to mass. 
The data support this scenario since the $\langle f_p \rangle$ 
distributions are much flatter 
than the $\langle f_{\pi^+} \rangle$ ones. 
The $m_T$ distribution of $\langle f_p \rangle$ 
becomes flatter as $\sqrt{s_{nn}}$ increases 
because of an increase in flow and/or freeze-out temperature. However
in general the velocity profile cannot be determined without knowing
the density profile and so a determination of a mean velocity from 
$\langle f_p \rangle$ is beyond the scope of this work, \cite{POLL98A}.

Several hydrodynamical models have interpreted the HBT source radii as 
``lengths of homogeneity'' which should decrease 
with increasing $m_T$ and this is consistent with
our data \cite{CSOR94A,WEID99A}.
Both pion and proton radii increase with $\sqrt{s_{nn}}$. However
$\langle f_{\pi^+} \rangle$
increases with $\sqrt{s_{nn}}$ while $\langle f_p \rangle$ drops.
Since $\bar p/p \ll 1$ at both SPS and AGS energies 
\cite{NA44pmp,E878,E864}, 
we know that most protons observed near mid-rapidity are remnants of 
the target or projectile that were 
slowed down by multiple 
collisions.
At the higher energy the protons occupy a somewhat larger  volume and
they are spread over a
larger momentum ({\it i.e.} $y,m_T$) range.
Therefore  $\langle f_p \rangle$ drops with $\sqrt{s_{nn}}$.

For pions the situation is different.
At $\sqrt{s_{nn}}$ = 4.9 GeV, they are outnumbered by protons and so
their freeze-out is driven by that of the nucleons. 
At $\sqrt{s_{nn}}$ = 17.3 GeV, they are the most numerous particle and
control freeze-out. 
Since ${\sigma}_{\pi\pi}<{\sigma}_{p\pi}$
$\langle f_{\pi^+} \rangle$  increases with $\sqrt{s_{nn}}$. Note that the ratio
$\langle f_{\pi^+} \rangle/\langle f_p \rangle$ increases by a factor of about
16 from  $\sqrt{s_{nn}}$ = 4.9 GeV to 17.3 GeV while the $\pi/p$ ratio only
increases by a factor of 7.

Using the (anti)deuteron as a measure of the nucleon-nucleon correlations we
have used a coalescence formalism to make
the first measurements of $p$ and $\bar p$ source radii and phase space
densities as a function of
$m_T$ at $\sqrt{s_{nn}}=17.3$ GeV.
At $\sqrt{s_{nn}}$ = 4.9 GeV the antiproton source is smaller than
the proton source while at $\sqrt{s_{nn}}=17.3$ GeV it appears to be larger.
%The antiproton source radius is larger than the proton source radius.
We have compared
the proton data to our $\pi^+$ radii
and phase space densities derived from HBT and single particle results 
as a function of system size and $\sqrt{s_{nn}}$. 
This comparison reveals a linkage between proton and pion freeze-out 
that changes  as the $\pi/p$ ratio increases.
At $\sqrt{s_{nn}}$ = 4.9 GeV the hadronic system is held together
by protons while at $\sqrt{s_{nn}}$ = 17.3 GeV it is held together by pions.

We thank
the NA44 collaboration for permission
to publish deuteron spectra and much valuable input. Thanks also to
 G. Bertsch, U. Heinz, S. Mr\'owczy\'nski   and S. Pratt for
helpful discussions.

\begin{thebibliography}{99}
\bibitem{PBM99A} P. Braun-Munzinger, I. Heppe and J. Stachel,
Phys. Lett. B {\bf 465} 15 (1999).
\bibitem{NA49ET} T. Alber {\it et al.}, Phys. Rev. Lett. {\bf 75} 3814 (1995).
\bibitem{LQCD} A. Ukawa, Nucl. Phys. A {\bf 638} 339c (1998).
\bibitem{NA44}
H. Beker {\it et al.},  Phys. Rev. Lett. {\bf 74}, 3340 (1995);
K. Kaimi {\it et al.},  Z. Phys. C {\bf 75} 619 (1997); 
H. B\o ggild {\it et al.}, Phys. Rev. C {\bf 58}, 328 (1999)  nucl-ex/9808002;
H. B\o ggild {\it et al.}, Phys. Lett. B {\bf 349} 386 (1995).
\bibitem{NA44pbk} I.G. Bearden {\it et al.}, Phys. Lett. B {\bf 471}, 6 (1999).
\bibitem{NA44pbpi} I.G. Bearden {\it et al.},  Phys. Rev. C {\bf 58}, 1656 (1998).
\bibitem{NA44spp} H. B\o ggild {\it et al.}, Phys. Lett. B {\bf 458}, 181 (1999).
\bibitem{NA44rvm} I.G. Bearden {\it et al.}, 
``Space-time evolution of the hadronic source in
peripheral to central Pb+Pb collisions." accepted by Eur. Phys. Jour. C May 2000.
\bibitem{NA44pmp} I.G. Bearden {\it et al.}, Phys. Rev.  C {\bf 57}, 837 (1998).
\bibitem{NA44dbar} I.G. Bearden {\it et al.}, 
Phys. Rev. Lett. {\bf 85} 2681 (2000).
\bibitem{SORGE95} H. Sorge,   Phys. Rev. C {\bf 52}, 3291 (1995).
\bibitem{thankN44} I would like to thank my NA44 collegues for agreeing  to
publication of  some of our
deuteron spectra before a comprehensive paper is published.
\bibitem{E866pd} L. Ahle {\it et al.},  Phys. Rev. C60 064901 (1999).
\bibitem{HANSENQM} A. Hansen {\it et al.}, Nucl. Phys. A. {\bf 661} 387c (1999).
\bibitem{BUTLER} S. Butler and C. Pearson,  Phys. Rev. {\bf 129}, 836 (1963).
\bibitem{KAPUSTA} J. Kapusta,  Phys. Rev. C {\bf 21}, 1301 (1980).
\bibitem{COALDATA} H. Gutbrod {\it et al.}, Phys. Rev. Lett. {\bf 37} 667 (1976);
S. Nagamiya, M. C. Lemaire, E. Moeller, S. Schnetzer, G. Shapiro, H. Steiner,
 and I. Tanihata, Phys. Rev. C {\bf 24} 971 (1981);   B. V. Jacak, D. Fox, and G. D. Westfall
{\it ibid.}  {\bf 31} 704 (1985);
J.W. Cronin {\it et al.}, Phys. Rev. D {\bf 11}, 3105 (1975).
\bibitem{MEKJIAN} A.Z. Mekjian,  Phys. Rev. C {\bf 17}, 1051 (1978).,
Nucl. Phys. A {\bf 312}, 491 (1978).  
\bibitem{E864Neut} T.A. Armstrong, {\it et al.}, Phys. Rev. C {\bf 60} 064903 (1999).
\bibitem{SCOTT95} W.J. Llope {\it et al.},  Phys. Rev. C {\bf 52}, 2004 (1995).
\bibitem{HODGSON} P.E. Hodgson, Nuclear Reactions and Nuclear Structure p453,
Clarendon Press, Oxford, (1971).
\bibitem{MROWCZ92} S. Mr\'owczy\'nski  Phys. Lett. B {\bf 277} 43 (1992).
\bibitem{SCHEIBL} R. Scheibl and U. Heinz,  Phys. Rev. C {\bf 59},
                  1585 (1999).
\bibitem{HEINZ96} U. Heinz, B. Tom\'a\v{s}ik, U.A. Wiedemann, and Y.F. Wu,  Phys. Lett. B {\bf 382}, 181 (1996).
\bibitem{CHAPMAN} S. Chapman, P. Scotto, and U. Heinz, Phys. Rev. Lett.
{\bf 74} 4400 (1995) and Heavy Ion Phys. {\bf 1} 1 (1995).
\bibitem{CSOR94A} T. Cs{\"o}rg\H o and B. L{\"o}rstad,  Phys. Rev. 
                  C {\bf 54} 1390 (1996).
\bibitem{WEID99A} U.A. Weidemann and U. Heinz, 
 Phys. Rep {\bf 319} 145 (1999)  nucl-th/9901094. %CERN-TH/99-15
\bibitem{POLL98A} A. Polleri, J.P. Bondorf, and I.N. Mishustin,
Phys. Lett. B {\bf 419} 19 (1998). 
\bibitem{CSOR96A} T. Cs{\"o}rg\H o, B. L{\"o}rstad, and
J. Zim\'{a}nyi Z. Phys. C {\bf 71} 491 (1996).
\bibitem{HUBERT} P. Avery {\it et al.}, Phys. Rev. D {\bf 32} 2294 (1985).
\bibitem{JOHNQM97} J. Sullivan, 
http://p25ext.lanl.gov/people/sullivan/talks\
/na44/hbt$\_$coul$\_$poster.html
\bibitem{BERTSCH} G.F. Bertsch, Phys. Rev. Lett. {\bf 72} 2349 (1994).; \\
{\it ibid.} {\bf 77} (1996) 789(E).
\bibitem{E87797} J. Barrette, {\it et al.},  Phys. Rev. Lett.  {\bf 78}, 2916 (1997),
Nucl. Phys. A {\bf 312}, 491 (1978). 
\bibitem{MURRAYQM} M. Murray {\it et al.}, Nucl. Phys. A {\bf 661} 456c (1999).
\bibitem{SORGE95B} H. Sorge, J.L. Nagle, and B.S. Kumar, Phys. Lett. B {\bf 355}, 27 (1995).
\bibitem{SINUKOV94} Y.M. Sinukov, Nucl. Phys. A {\bf 566f} 589c (1994).
\bibitem{NA52dbar} G. Ambrosini {\it et al.},  CERN-OPEN-99-309,
New J. of Phys. {\bf 1} 22 (1999).
\bibitem{NA49pbpp} H. Appelsh{\"a}user {\it et al.},  Phys. Lett. B {\bf 467}, 
21 (1999) nucl-ex/9905001.
\bibitem{FERENC} D. Ferenc, U. Heinz, B. Tom\'a\v{s}ik, U.A. Wiedemann, and J.G. Cramer,
  Phys. Lett. B {\bf 457}, 347 (1999) hep-ph/9902342.
\bibitem{MROWCZ93} S. Mr\'owczy\'nski  Phys. Lett. B {\bf 308} 216 (1993).
\bibitem{E864dbar} T.A. Armstrong {\it et al.}, Phys. Rev. Lett. {\bf 85} 2685 (2000), nucl-ex/0005001
\bibitem{JACAK95} B. V. Jacak {\it et al.}, Nucl. Phys. A {\bf 590} 
215c (1995).
\bibitem{TOMASIK} B. Tom\'a\v{s}ik, U.A. Wiedemann, and U. Heinz, nucl-th/9907096.
\bibitem{E866jh} R. A. Solz {\it et al.}, Nucl. Phys. A {\bf 661} 439c (1999).
\bibitem{E802AHLE} L. Ahle {\it et al.},  Phys. Rev. C {\bf 57}, R466 (1998).
\bibitem{E878} M.J. Bennett {\it et al.},  Phys. Rev. C {\bf 56}, 1521, (1997).
\bibitem{E864} T.A. Armstrong {\it et al.},  Phys. Rev. C {\bf 59}, 2699, (1999).
\end{thebibliography}
\end{document}